\documentclass{article}
\usepackage[english]{babel}

\usepackage[letterpaper,top=2cm,bottom=2cm,left=3cm,right=3cm,marginparwidth=1.75cm]{geometry}

\newcommand\be{\begin{equation}}
\newcommand\ee{\end{equation}}
\newcommand\ben{\begin{equation*}}
\newcommand\een{\end{equation*}}
\newcommand\ba{\begin{eqnarray}}    
\newcommand\ea{\end{eqnarray}}      
\usepackage{amsmath}
\usepackage{amssymb}
\usepackage{xcolor}
\usepackage{graphicx}
\usepackage[colorlinks=true, allcolors=blue]{hyperref}

\usepackage{graphicx} 

\usepackage[none]{hyphenat} 
\numberwithin{equation}{section} 
\usepackage{caption}
\usepackage{subcaption}
\usepackage{float}

\title{The (No) Boundary Proposal and \\

excited states in de Sitter holography}
\author{ }
\date{ }

\begin{document}
\begin{titlepage}

\maketitle

\begin{center}

\vspace{-1cm}

Marcelo Botta-Cantcheff${}^{\dagger}$, Facundo Lorenzo Cruz${}^{*}$ and Pedro J. Martínez${}^{\ddagger}$

\vspace{.8cm}

\small{  Instituto de Física La Plata, IFLP-CONICET}

\vskip .15cm

\small{\textit{Diagonal 113 e/ 63 y 64, La Plata, Argentina}}

\vskip 2cm

\begin{abstract}

In the AdS/CFT framework, vacuum and excited states are systematically described by imposing arbitrary Dirichlet boundary conditions at the AdS boundary. Furthermore, there are explicit relations connecting the quantum states to their corresponding dual Euclidean AdS geometries, in line with the Hartle-Hawking (HH) construction.
The ground state therefore corresponds to the dominant saddle  point under trivial conditions on the asymptotic boundary, which is the exact Euclidean AdS geometry.
 In contrast, the situation in de Sitter spacetime differs significantly, as there is no natural region analogous to the AdS boundary.
Thus, the Hartle Hawking approach precisely defines the ground state as a path integral over smooth (Euclidean) geometries ending on a spatial Cauchy surface, with \textit{no} additional  boundary or past singularity, known as the no boundary proposal.

In this work, we revisit the no boundary proposal to describe excited states within the framework of de Sitter Holography. Specifically, we investigate the possibility of defining a family of excited states by introducing an additional boundary in the Euclidean region and imposing arbitrary Dirichlet boundary conditions on it. As a result, we demonstrate that the computation of $n$ point correlation functions is consistent with the presence of excited states, and furthermore, show that cosmological late-time observables in these states undergo non-trivial modifications. This study may have significant implications for the development of the holographic dictionary for de Sitter spacetimes.

\end{abstract}
\end{center}

\small{\vspace{6.5 cm}\noindent ${}^{\dagger}$botta@fisica.unlp.edu.ar\\\noindent ${}^{*}$facundo.cruz@fisica.unlp.edu.ar \\
${ }^{\ddagger}$martinezp@fisica.unlp.edu.ar}

\thispagestyle{empty}

\end{titlepage}

\setcounter{tocdepth}{2}

{\parskip = .4\baselineskip \tableofcontents}

\section{Introduction}

\begin{figure}[t]\centering
\includegraphics[width=.9\linewidth] {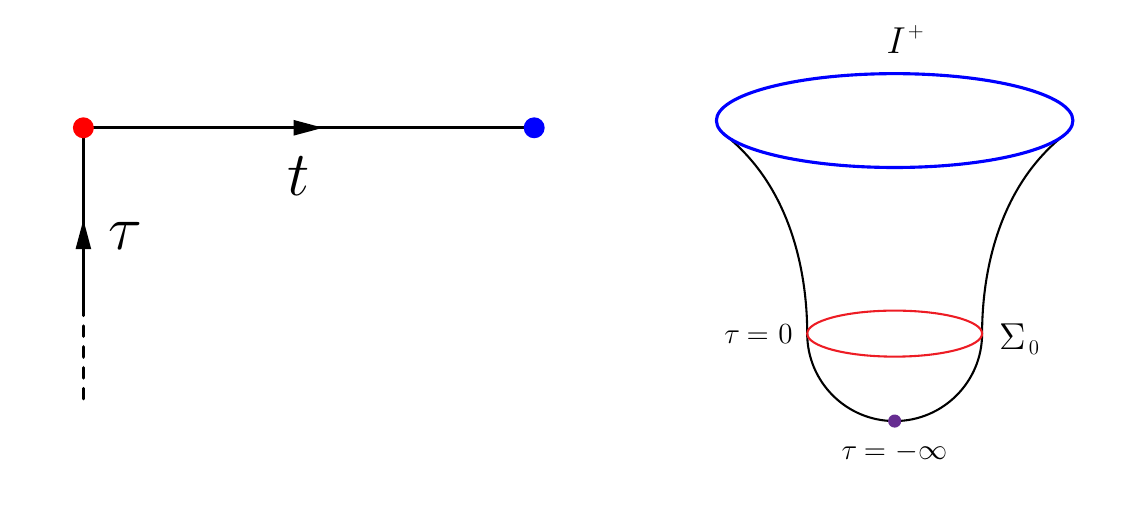}
\caption{\small{de Sitter spacetime is represented as the real-time evolution from the vacuum state at $\tau=t=0$. This is described as the lower half of Euclidean dS space.The corresponding Schwinger-Keldysh diagram in the complex plane is shown on the left.}}
\label{NBP1} 
\end{figure}

In the path-integral formulation of ordinary quantum mechanics and Quantum Field Theory (QFT), the analytic continuation to imaginary times provides a description of the vacuum state. This depicts an interpretation of Euclidean geometries in terms of quantum states, particularly useful for quantum gravity and in the cosmological context, assuming that the universe began in its ground state. This is known as the Hartle-Hawking construction \cite{HH}, and in the inflationary model where spacetime is described by de Sitter geometry, it implies that at the onset of expansion, when quantum gravitational effects must be considered, the geometry can effectively be continued to Euclidean de Sitter (see Fig. \ref{NBP1}), which has no boundaries and is often referred to as the Euclidean vacuum. At a semi-classical level, the contribution of this geometry to the action fits the wave function of the ground state, providing the amplitudes for a certain initial configuration of all the fields and the spatial geometry, $\phi = \Phi|_\Sigma$. The analytically continued geometry has no boundary other than the Cauchy hypersurface, $\Sigma$, where the real time evolution begins. Even when generalized to include the effects of gravity, the wave function $\Psi_{dS} [\phi]$ is defined as a sum over Riemannian geometries with compact topologies, that is, without boundaries \cite{HH}.
This is the No Boundary Proposal (NBP), and it is particularly useful to formulate the holographic recipe in the context of de Sitter spacetimes accurately, which is the focus of the present study.

On the other hand, the well-established AdS/CFT correspondence has led to the natural hypothesis that this can be generalized to other asymptotic behaviors of spacetime; thus, it is claimed that spacetimes that for very late times behave as de Sitter, can be equivalently described by a conformal theory defined on the future asymptotic boundary $I^+$ \cite{Strominger2001}.
Because the isometries of the de Sitter spacetime in $d+1$ dimensions are $SO(1,d+1)$, the dual theory must have the same conformal symmetry group, therefore it must be a theory defined on a Euclidean space of dimension $d$, identified with $I^+$.
Moreover, matching between $n$-point functions computed for a field theory on $dS_{d+1}$ 
 at very late times, and correlators between primaries of a CFT$_d$, constitutes a strong evidence that in fact quantum fields on de Sitter behave as CFT$_d$, at least kinematically \cite{Strominger2001,Dio-Stro2011,Galante23}.
 
The most accepted prescription that implements this proposal and captures the behavior of the correlation functions in both theories is \cite{Malda03, Bauman}
\be\label{recipe-vac} \Psi_{dS} [\phi] = Z_{CFT} [J] \ee
This establishes the correspondence between both theories $dS_{d+1}/$CFT$_d$ and explicitly involves the Hartle-Hawking state, which can be computed by virtue of the NBP. In fact, $\Psi_{dS} [\phi] $ stands for the wave functional of the fundamental state for (all) the bulk fields defined on a de Sitter geometry, described by the (lower) half of the Euclidean dS (Fig. \ref{NBP1}), and then evolved to very late times near the asymptotic boundary $I^+$. The $Z_{CFT} [J]$ on the right hand side stands for the generating function of the CFT defined on the Riemmanian $d$-sphere, 
whose points ${\bf x}$ can be identified with $I^+$, and thus $J({\bf x})$ is an arbitrary source for the primary operator ${\cal O}$, considered dual to the bulk field $\Phi$.

This formula allows to compute all the In-In diagrams in dS, and relate them to correlation functions in the CFT \cite{Higuchi:2010xt, Bauman}. To do it one must identify the variations 
\be
\label{phiJ}\delta \phi({\bf x}) = \delta J ({\bf x})\;.
\ee
However, many aspects of this holographic dictionary remain to be fully clarified, for example, how to describe states other than the Euclidean vacuum of dS and how these should be related to the CFT, see e.g. \cite{Doi:2024nty} for a recent attempt in low dimensional theories. This is the main objective of our study and for this we propose to generalize the NBP in the simplest possible way: by imposing, by hand, a boundary in the Euclidean region with Dirichlet boundary conditions.

It is not even known actually if there can be more states in de Sitter than the (Euclidean) dS vacuum \cite{Higuchi1,Higuchi2}, or if any consistent attempt at generalization captures or not other types of ground states, such as those described in the literature as $\alpha$-vacua \cite{Allen}, see also e.g. \cite{deBoer:2004nd}. There have been other approaches to the problem of defining excited states in dS \cite{excited1,excited2}, particularly in the context of holography, which have been previously discussed in the literature, see for instance \cite{Betzios:2024oli}. 
Our approach is nothing but than an attempt to build states other than the (Euclidean) dS vacuum, in a way that it be consistent with the dual CFT through a suitable generalization of this formula, and with the basic requirements that can be expected from a holographic point of view.

Finally, one would like to be able to describe states that, in the same
sense as in AdS, are dual to some kind of classical geometry that approaches to the de Sitter
space-time near the asymptotic future region.
The approach proposed here enjoys this property by construction, since the states are built from the gravitational path integral.

The manuscript is organized as follows. In Section 2, we review the NBP-based construction leading to the standard holographic formulation for dS/CFT; therefore, we propose a generalization that is capable of consistently capturing initial states more general than the vacuum state. In Section 3, we study the  implementation this framework to the $dS_{1+1}$ case, and present the detailed computation of the correlation functions regarding the new prescription.
Concluding remarks and some open questions for future research are collected in Section 4. Finally, the article has two appendices that describe some relevant aspects and technical details involved in the work.

\section{On the NBP and its generalization}

NBP has been characterized as a theory to prescribe the initial conditions on the hypersurface $\Sigma_0$ \cite{malda24}. In the same spirit, the proposal we aim to explore here is heuristically supported by observations related to the original Hartle-Hawking construction and holography.

Firstly, it is well-known that for a quantum field theory (QFT) defined on a static spacetime $M$, where the Hamiltonian $H$ is time-independent, and along the timelike Killing vector's affine coordinate, one can perform the standard analytic continuation of the time coordinate $\tau = -it$ to values $-\infty < \tau \leq 0$. The state of the system evolves according to 
$$
|\psi(\tau=0) \rangle = e^{H\tau} |\psi(\tau) \rangle.
$$
It is important to note that if $H$ is Hermitian and bounded from below, its eigenvalues satisfy $E \geq E_0 \geq 0$, then this state approaches the vacuum $|E_0\rangle$ in the limit as $\tau \to -\infty$. This property is independent of the chosen ``initial'' state $|\psi(-\infty) \rangle$, which only influences a normalization factor. In other words, the evolution operator $e^{H(\tau \to -\infty)}$ effectively acts as a projector onto the vacuum state. A key observation for our purposes is that if one does not take this limit and instead evolves from a finite Euclidean time $\tau_b < 0$, the resulting state at $t = \tau = 0$ is a quantum superposition of excited states with different energies. The amplitudes of this superposition depend on the configuration at $\tau_b$. In the context of Hartle-Hawking’s construction, $\tau_b$ represents a boundary for the Euclidean region of spacetime. Our proposal seeks to generalize the NBP by suggesting a similar boundary in de Sitter spacetime to construct new states with a holographic interpretation.

It is important to note that this argument serves to heuristically justify the proposal, since de Sitter spacetime is not static, and $H$ is generically time-dependent. Nevertheless, we will investigate the possibility of imposing arbitrary Dirichlet boundary conditions at the additional boundary $\tau_b$.

In the context of AdS holography, there exists a construction that aligns with this approach, successfully describing the vacuum and its excitations by prescribing Dirichlet conditions on the asymptotic component of the Euclidean boundary, which is connected to the hypersurface $\Sigma$ at $\tau=0$ \cite{SvR1,SvR2, us1}. This further reinforces the proposal.

In order to circumvent theoretical and technical issues associated with gravitational constraints, we exclude gravity from the current study and assume low-energy field fluctuations, ensuring that back-reaction effects remain negligible.
However, it is important to mention that recent studies have explored the conditions under which geometries such as the one considered here can appear as gravitational saddle \cite{Friedrich:2024aad}.

\subsection{The holographic formula}

Recipe \eqref{recipe-vac} is often defined in a Poincaré coordinate system, where a quantum field theory is well defined and can be canonically quantized and treated perturbatively, e.g. \cite{Malda03, Bauman}, and in fact, the canonical (Bunch-Davies) vacuum computationally coincides with the wave functional on the RHS.

For our goals on this note, we would like to present the prescription \eqref{recipe-vac} in a more convenient form, which is manifestly independent of the coordinates frame. 
For simplicity, let us consider here a global foliation of the  de Sitter \textit{expanding} spacetime in Cauchy surfaces: $M = \{ \Sigma_t  \, , \,  0\leq t\leq T \}$, where $\Sigma_t$ are compact $d$-spheres.

So, the left hand side of \eqref{recipe-vac} stands for the probability amplitude of a specific configuration $\phi_T({\bf x})$ at the late time $T$ spacelike hypersurface $\Sigma\equiv \Sigma_T$, evolved in the real time from the fundamental state prepared at $t=0$ in real time :

\be\label{PsiU} \Psi_{dS}  \equiv \langle \phi \equiv \phi({\bf x}), T | \, {\cal P}_{\{t\}} \,e^{-i\int^T_{0} d t\; H(t)} \,|0 \rangle \,\ee
where $|0\rangle$ represents the lower (Euclidean) part of the geometry shown in Fig. \ref{NBP1}.
This can be expressed by a standard Feynman path integral on a de Sitter spacetime with \textit{Lorentzian} ($L$) signature
$$ Z^{L}_{dS}(\phi, \phi_0) = \int^{\Phi|_{T}= \phi({\bf x})}_{\Phi|_{0}= \phi_0({\bf x})} [\mathcal{D}\Phi] \,  e^{i S^{}[\Phi]}\,\,,$$
but it depends on the initial configuration of fields $\phi_0({\bf x})$ on $ \Sigma_0$, whose probability amplitude is $ \langle \phi_0({\bf x})| \,0 \rangle \equiv \Psi(\phi_0) $. Precisely, the NBP provides this completely, by analytically continuing into the Euclidean de Sitter solution, in such a way that it does not requires any other boundary data.

\begin{figure}[t]\centering
\includegraphics[width=.9\linewidth] {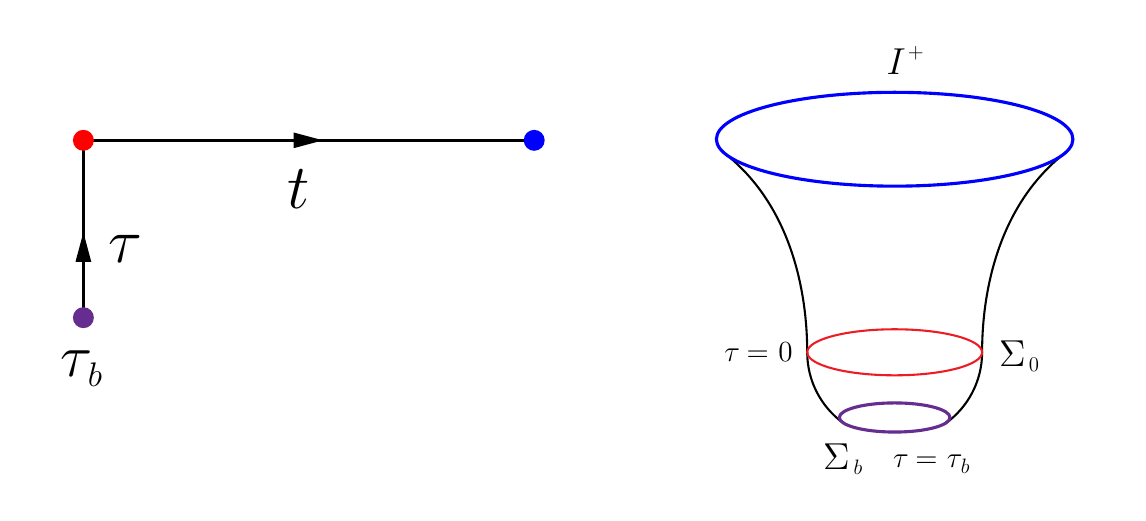}
\caption{\small{ The left figure is the Schwinger-Keldysh complex diagram, where the initial state is described as a path integral on a finite imaginary time interval $(\tau_b, 0)$. The right figure represents the corresponding saddle point geometry.} }
\label{NBP2}
\end{figure}

In fact, by applying the identity operator $I \equiv \int [d\phi]_{\Sigma_0} \;|\phi\rangle  \langle \phi | \,$ expressed in the field configuration basis on the vacuum state in eq. \eqref{PsiU}, we can express it as 
$$  Z^{L+E}_{dS}(\phi) = \int [d\phi']_{\Sigma_0} \; Z^{Lorentz}_{dS}(\phi, \phi')\, \Psi(\phi'({\bf x})) $$
Most notably, the rightmost factor is nothing but the wave-functional of the fundamental state, which according to the HH formalism can be written in Euclidean path integral form as,
$$\Psi(\phi) \equiv \int_{ \Phi|_{\Sigma_0} = \phi } [\mathcal{D}\Phi] \,  e^{-S^{E}[\Phi]}\,\, ,$$
which by virtue of NBP only requires one boundary data on $\Sigma_0$. In fact, it is a sum over fields $\Phi(\textbf{x}, \tau)$ on the Euclidean part of (compact) geometry $M_E$, whose only boundary\footnote{When gravity is included in the analysis one must also sum over Euclidean geometries anchored by the common boundary $\Sigma_0$, see e.g. \cite{SvR1,SvR2}.} is $\Sigma_0$, then it is glued to $M$ at $t=0$ , see Fig. \ref{NBP2}.
In such a sense, one can actually express the recipe as equality between two path integrals.
\begin{equation}\label{E+L-Correspondence}
Z^{L+E}_{dS}(\phi)= Z_{\text{CFT}} \,,
\end{equation}
where the path integral on the left sums over all the fields on a de Sitter geometry with signature change at $\Sigma_0$, as drawn in Fig. \ref{NBP1}. This is normally computed in saddle point approximation, which imposes smoothness of the field on this surface 
\be\label{gluing}  \Phi({\bf x}, t) |_{t=0^+} = \Phi({\bf x}, \tau) |_{\tau=0^-}\,~,~~\partial_t \Phi({\bf x}, t) |_{t=0^+} = -i \partial_\tau \Phi({\bf x}, \tau) |_{\tau=0^-}\ee
and determines $\phi_0$ uniquely on terms of the b.c. $\phi$.
Moreover, the classical e.o.m. is elliptic, so then, the solution $\Phi({\bf x}, \tau)$ in the interior of $M_E$ is also uniquely determined by this Dirichlet data. 
For simplicity, we have assumed here a free (scalar) field, but more general theories can be considered and the gluing conditions are derived within the formalism \cite{SvR2}.
If gravity is taken into account, the Israel junction conditions arise \cite{Israel}.

Thereby, in the semiclassical approximation we finally have
\be\label{wf0-onshell}\Psi(\phi) = e^{iS^{L}(\phi,\phi_0)} \,  e^{-S^{E}(\phi_0)}\,\, ,\ee
where $S^{L/E}$ denote the on shell action, as function of their respective boundary conditions. The solutions of the e.o.m. are solved on the manifold of Fig. \ref{NBP1}, $M\cup M_E$ and $\phi_0$ is determined by conditions \eqref{gluing}.

As previously discussed, this demonstrates that NBP univocally determines the initial condition $\phi_0$, which is closely linked to the fundamental state, eq. \eqref{PsiU}. Thus, in order to describe a more general set of states such that this configuration changes $\phi_0 \to \phi'_0$, the simplest and most natural proposal is to relax this requirement by arbitrarily imposing a new boundary condition in the Euclidean region that will characterize the state.

Thus, let us introduce  a compact boundary at $\Sigma_b$ as seen in Fig. \ref{NBP2} 
\be\label{PsiU-SK-b}\Psi'(\phi'_0 , \phi_b) \equiv \int_{ \Phi|_{\Sigma_{0.b}} = \phi'_{0}, \phi_{b} }~ [\mathcal{D}\Phi]_{M_E^b} \,~~  e^{-S^{E}[\Phi]}\,\, , \ee
where $M_E^b$ is an Euclidean compact space with two disconnected (compact) boundaries $\Sigma_{0} , \Sigma_{b}$ and  the sum is over all the field configurations on it. In the foliation analytically continued $M_E^b =\{ \Sigma_\tau  \, , \,  \tau_b \leq \tau \leq 0 \}$, this state is nothing but 
\be\label{PsiU-b} \Psi'_{dS}  \equiv \langle \phi'_0 , 0 | \, {\cal P}_{\{\tau\}} \,e^{\int^0_{\tau_b} d \tau\; H(\tau)} \,|\phi_b , \tau_{b}\rangle \,\ee
therefore, by composing this with \eqref{PsiU}, the new wave functional results from evolving the field configuration $\phi_b, \Sigma_b$ by a (path ordered) complex Schwinger-Kelldysh curve depicted on the left of Fig. \ref{NBP2}
\be\label{PsiU-SK-b-Ope} \Psi'_{\phi_b} (\phi, T) \equiv \langle \phi , T | \, {\cal P}_{\{s\}} \,e^{\int^T_{\tau_b} d s\; H(s)} \,|\phi_b , \tau_{b}\rangle \,\ee
where $s$ parametrizes the Schwinger-Keldysh curve in the complex plane. Our claim is that \eqref{PsiU-b} characterizes the state $\phi_b, \Sigma_b$. 
So we can write the more general holographic prescription as follows
\be\label{recipe-stateb} \Psi'_{\phi_b} (\phi, T) = Z_{\text{CFT}}(J) \,\ee

It is worth noticing that as expected, states defined through path integrals in this way, correspond to good geometries and field configurations in the semiclasscial limit. Then in addition, they will have the corresponding CFT dual as we will show below. 

We must us verify that in fact this new proposal in fact fits into the expected feature of a holographic prescription for excited states, as discussed in the forthcoming subsection.
In what follows we study these initial conditions and their consequences. Moreover, we shall investigate more the role played by the sources $\phi$, and  $J$, in this formula.

\subsection{Consistency with CFT excitations}

The right-hand side of equation \eqref{recipe-vac} is a generating function associated with a generic source $J({\bf x})$, which plays an auxiliary role in the formula for obtaining the series expansion in perturbations $\delta J$. Through this expansion, one can recover the correlation functions of the source-free CFT model, which are ultimately related to the components of the wave functional $\Psi_{dS}$ \cite{Bauman}. In fact,

\be\label{CFTstate} Z_{\text{CFT}}(J) = \int[D\eta] \; e^{- S_{\text{CFT}}[\eta]\, + \,\int_{S^d} J\, {\cal O}} \;,\ee
where $\eta({\bf x})$ denotes the CFT fields defined on $S^d$. Thereby, the generalized formula \eqref{recipe-stateb} implies that this expression, or its interpretation, shall be consistently modified.

On the other hand, let us recall that excited states of a CFT$_d$ can be systematically described by a source term $\int J {\cal O}~$\,, added to the action with certain non vanishing source  $J$ as above. This has been extensively studied and confirmed in several ways, and particularly in the context of the AdS/CFT correspondence \cite{us1,us3,kaplan}.
In fact, according to the operator-state map, excited states are constructed from primary operators inserted at specific points ${\bf x}_j$ of $S^d$, so for instance, the simplest excited state ${\cal O}({\bf x}_j)|0\rangle$ can be expressed through a source $J({\bf x}) = \lambda \delta({\bf x}-{\bf x}_j)$, by considering a series expansion of \eqref{CFTstate} in the real parameter $\lambda$.

Therefore, in what follows we will define excited states in the dual CFT by imposing some generic external source $J_b$. So we must verify that the new proposed formula \eqref{recipe-stateb} can be in fact consistent with it, and what are its implications.

The first consequence of this is that one point function of primary operators in such state does not vanish for all ${\bf x}$ \be \label{expectedO}\langle{\cal O}({\bf x})\rangle_{J_b} \neq 0 \ee

Notice that for the vacuum, the on-shell wave functional \eqref{wf0-onshell} for a scalar real field reads
\be\label{wf0-onshell-gen}\Psi(\phi)= N \exp \left[ \sum_{n=2}^\infty\,  \frac{1}{n!}  \int_{\Sigma_T} d{\bf x}_1 \dots d{\bf x}_n~  \psi_n({\bf x}_1 ,\dots , {\bf x}_n) \, \phi({\bf x}_1) \dots \phi({\bf x}_n) \right].\ee
For a free theory it is a Gaussian functional with only the coefficient $\psi_2 \neq 0$, which can be obtained by exactly solving the path integral, and the $n>2$ terms come from the interaction terms. By virtue of the expected duality one can relate the non-trivial coefficients of this expansion with CFT computations.

Taking functional derivatives to both sides of the formula \eqref{recipe-vac}, using \eqref{CFTstate} with the variations are related by \eqref{phiJ}, we obtain that the coefficients are given by $n$-point correlation functions of the sourceless CFT:
 \be\label{On-psi}  \psi_n =  ~  \langle{\cal O}({\bf x}_1) \dots {\cal O}({\bf x}_n) \rangle_{J=0}  ~~.\ee
Moreover,  $  \psi_1 =0$, is consistent with the fact that $$ \langle{\cal O}({\bf x})  \rangle_{J=0} =0  .$$ 

Thereby, the simplest check that states built with b.c. $(\Sigma_b, \phi_b)$, can be consistent with dual CFT excitations such that the formula \eqref{recipe-stateb} is preserved, is the following.

Having into account up to $n=2$ for simplicity, the state \eqref{PsiU-SK-b} can be expressed in terms of its e.o.m. solution
 \be\label{logPsiU-SK-b} \log \Psi'[\phi , \phi_b] = \,~ \int_{\Sigma_T}d{\bf x} \;\int_{\Sigma_b} d\textbf{y} ~\phi({\bf x}) ~ K_{b}({\bf x},{\bf y}) ~ \phi_b({\bf y})~ + ~\dots\, , \ee
 here $\dots$ expresses terms that are bilinear in $\phi$ and $\phi_b$ respectively, which are not important now. This is the term that precisely provides a non vanishing one point function
 \be\label{1pt-exc1}\,\int_{\Sigma_b} d{\bf y}  ~ K_{b}({\bf x},{\bf y}) ~ \phi_b({\bf y})~ = ~ \langle{\cal O}({\bf x})\rangle_{J_b} ~,\ee
which is consistent with \eqref{expectedO} as expected. Remarkably, it also implies that in presence of the boundary $\Sigma_b$, we must have
$$J({\bf x}) = J_b({\bf x}) + \delta J({\bf x}) $$ in the generating function \eqref{CFTstate}, for some $J_b({\bf x}) \neq 0$.
 Thus, in what follows this feature will be shortly referred to as excited states.

It is worth emphasizing that within the present proposal, both the presence of $\Sigma_b$ and the non vanishing source $\phi_b \neq 0$ contribute to deforming the Euclidean vacuum. Consequently, taking the limit $\phi_b \to 0$ does not necessarily restore the correlators of the undeformed vacuum, as the geometry still contains a hole and a residual dependence on $\tau_b$ persists. In Section \ref{BPCorrelators} we will show an example (dS$_{1+1}$) that explicitly realizes this  framework.

\subsection{Modified late time correlators in excited states}

The novel wave functional of states \eqref{PsiU-SK-b-Ope} for a theory of a real scalar field above reads
\begin{equation}\label{Exc-WF}
     \Psi' = N' \,\exp\,\left\{ \,\int_{}dk^d \; \psi_1(k, J_b) \, \phi(k) ~ + ~\sum_{n=2}^\infty\, \frac{1}{n!}\,\int_{} \,\psi_n \, \phi^n \right\}~~
\end{equation}
expressed in the momentum space, where
 $\phi^n $ denotes $\phi(k_1) \dots \phi(k_n)$, and $\psi_n(k_i, J_b)$ are the coefficients corrected because the new boundary condition. Note the new term $n=1$, that we wrote separately, and that a non trivial $\psi_0$ term would contribute only to the normalization factor $N'$. Thus, from the prescription \eqref{recipe-stateb}, it can be shown that the wave functional coefficients are given by $n$-point functions of CFT operators
\be\label{On-psiexc}  \psi_n (k_i, J_b)  ~ =  ~ \langle{\cal O}(k_1) \dots {\cal O}(k_n) \rangle_{J_b} ~~,\ee
 and in particular \be\label{1pts-exc1}\,\psi_1 (k, J_b)=  ~ \langle{\cal O}(k) \rangle_{J_b}~.\ee
 It is worth noting that one must still take variations on both sides of \eqref{recipe-stateb} according to \eqref{phiJ}, but ultimately evaluate at $\phi=0$ on the left, and $J = J_b $ on the right.

 By virtue of the translational invariance we have $\psi_{n>1} (k_i, J_b) =\hat \psi_n \,\delta(\sum_{i=1}^n k_i)$, which expresses the momentum conservation. In a global foliation the Cauchy's surfaces are compact, and the numbers $k$'s are discrete.
The same computation can be done in a non compact foliation as the Poincare coordinates system, where $k_i \in \mathbb{R}^d$ are the Fourier modes. 

Late time $\phi$ correlators are related to $\cal O$'s $n$ point functions by means of the wavefunctional \eqref{Exc-WF} as
\begin{equation}\label{phi-npts}
    \langle\; \prod_{i=1}^n \phi(k_i)\;\rangle_{J_b}\equiv \frac{\int {\cal D}\Phi \prod_{i=1}^n \phi(k_i) |\Psi'|^2}{\int {\cal D}\Phi \; |\Psi'|^2 }~ ,
\end{equation}
and then using the relations \eqref{On-psiexc}. For a real scalar free field this is a Gaussian model and can be exactly computed, using that  
$\phi(k)=\phi(-k)^*$
we get
 \be\label{1pt-exc2}\langle \phi(k) \rangle_{J_b} = -\,\frac{\psi_1(k)+\psi_1^*(-k)}{2\Re e\{\hat\psi_2(k)\}}\ee
 \be\langle{\phi (k_1) \phi(k_2)}\rangle_{J_b}  = - \frac{\delta(k_1 + k_2)}{2\Re e\{\hat\psi_2(k_1)\}}+ \langle \phi(k_1) \rangle_{J_b} \langle \phi(k_2) \rangle_{J_b}\ee
where was used that $\psi_2 \equiv\hat \psi_2 (k_1) \, \delta(k_1 + k_2)$ to simplify the form of the 2 point function. Finally, one must substitute by \eqref{Exc-WF} to express it in terms of the operators ${\cal O}$, 
so in particular
\begin{equation}\label{delta-2pf}
       \hat \psi_2 (k, J_b)   \; =  \,  \langle{\cal O}(k) {\cal O}(-k) \rangle_{J_b} ~ ,
\end{equation}
that therefore, reduce to the known relations for $J_b =0$.

Note that these relations, valid for a free theory, agree with those obtained for the vacuum state by making the substitution:
\begin{equation}
   \phi'  (k) \equiv \phi (k) - \langle \phi_k \rangle ,
    \label{shift}
\end{equation}
which turns the functional wave back into a Gaussian.
However it cannot be interpreted as a meaningful redefinition of fields, because in fact, the background term (eqs. \eqref{1pt-exc1} and  \eqref{1pt-exc2}) implies that the field $\phi'(x)$ should be defined nonlocally and in a rather state-dependent way.

Inflation predicts that primordial correlations follow an approximately Gaussian distribution. However, the details of interactions during inflation are crucial, and in principle, could be detected in higher-order correlations associated with small deviations from Gaussianity \cite{Malda03, Bauman}.
For instance, let us consider the simplest model with a cubic interaction, described by a slight deformation in a coefficient $\psi_3\neq 0$. If the initial state is excited as described above, the holographic expression for the three-point late time correlation gets corrected, to leading order in $\psi_1$,
\be
   \langle \phi_1 \phi_2 \phi_3  \rangle_{J_b} =\langle \phi_1 \phi_2 \phi_3  \rangle_{J_b=0}+
  {\cal P} +  O^2 ( \psi_1(k))\label{3pfBoots}
\ee
where 
\begin{equation}\nonumber
    \langle \phi_1 \phi_2 \phi_3  \rangle_{J_b=0}=\frac{2 \Re e[\hat \psi_3(k_1, k_2, k_3)] \delta(k_1+k_2+k_3)}{
   \prod_{i=1}^3 (-2\Re e\{\hat\psi_2(k_i)\})}
\end{equation}
and
\be\nonumber
 {\cal P} =    \frac{\psi_1(k_1)+\psi_1^*(-k_1)}{2\Re e\{\hat\psi_2(k_1)\}} \frac{\delta(k_2 + k_3)} {2\Re e\{\hat\psi_2(k_2)\}}
  +{\rm perm.}\,_{(1\,2\,3)}
\ee
where we have neglected higher orders in $\psi_1(k)$ considering small deviations from the vacuum state,  
$|\psi_1(k)|\ll |\hat \psi_2(k)|$,
consistently with ignoring the back reaction effects, as well as a perturbative expansion in $\psi_3$. The one- and two-point functions also receive the corresponding corrections.

\section{Computing Correlators in dS with boundary}\label{BPCorrelators}
 
In this section we will consider our boundary proposal to define excited states and 
compute holographic and late time correlators in the simplest case of asymptotically $dS_{1+1}$. The set up is a massive scalar field in the Complementary representation, i.e. $m^{2}=\Delta(\Delta-d)$ with $\Delta\in[0,d]$ with $d=1$ in Global Coordinates.
To be concrete, we will choose $\Delta\in[0,d/2]$ keeping $\Delta<(d-\Delta)$ without loss of generality. We will often compare our results in this section against the No Boundary Proposal (NBP) results, which for self-consistency of the present work, we include in App. \ref{NBPCorrelators}.

We study $Z_{dS}^{L+E}$ over the manifold defined in Fig. \ref{NBP2}, foliating the manifold as
\begin{equation}\label{Euclid-Metric-BP}
ds_{dS}^2=\frac{d\tau^2+d\varphi^2}{\cosh(\tau)^2}\qquad\qquad\tau\in[\tau_b,0] \qquad\varphi\in[0,2\pi)
\end{equation}
We stress that the south pole is absent in our geometry. As in App. \ref{NBPCorrelators} we go to the Lorentzian signature piece as $\tau\to -i t$ and the Lorenztian piece of the geometry is the same as for Fig. \ref{NBP1}, as well as the boundary metric $ds^2_{CFT}=d\varphi^2$, see \eqref{B-Metric-NBP}. The Euclidean bare scalar field action defined as, 
\begin{align}
S_E=\frac 12 \int\!\! \sqrt{g} \left(\partial_\mu\Phi\partial^\mu\Phi+m^2\Phi^2\right)
\end{align}

The on-shell action of the scalar of the geometry in Fig. \ref{NBP2} has 2 pieces on top of counter-term found in \eqref{CT-Action}, i.e.
\begin{equation}\label{Sfin}
S_{\rm ren}=S_T+S_{b}-S_{\rm ct}=+\frac i2 \int_T\!\! \sqrt{\gamma} \;\Phi \,\partial_n \Phi-\frac 12 \int_{b}\!\! \sqrt{\gamma} \;\Phi \,\partial_n \Phi-S_{\rm ct}
\end{equation}
where $\Phi$ is the solution that meets,
\begin{equation}\label{EDM+BC}
\left(\square-m^2\right)\Phi=0\qquad\qquad\Phi_{n}(\pi/2-\epsilon)= \epsilon^{\Delta} \phi_n \qquad\qquad 
\Phi_{n}(\tau_b)=\phi_n^b
\end{equation}
where we have already introduced the asymptotic boundary source renormalization found in \eqref{Source-Renorm}, have defined late time at $T=\pi/2-\epsilon$ with $\epsilon\ll1$ and where the Fourier modes of the classical solution
\begin{equation}
    \Phi_{n}(\tau)=\int_0^{2\pi} \frac{d\varphi}{2\pi} e^{-in\varphi}\Phi(\tau,\varphi)
\end{equation}
are of the form
\begin{equation}
\Phi^E_{n}(\tau)=A_n P_{\Delta -1}^{|n|}(-\tanh (\tau ))+B_n P_{\Delta -1}^{|n|}(+\tanh (\tau ))
\end{equation}
\begin{equation}
\Phi^L_{n}(t)=A_n P_{\Delta -1}^{|n|}(+i\tan (t ))+B_n P_{\Delta -1}^{|n|}(-i\tan (t ))
\end{equation}
where the parameters $\{A_n,B_n\}$ are fixed in terms of $\{\phi_n,\phi^b_n\}$ by imposing \eqref{EDM+BC},
yielding
\begin{equation}
A_n=\frac{\epsilon^{\Delta} \phi_n P_{\Delta -1}^{|n|}(+\tanh (\tau_b )) - \phi^b_n P_{\Delta -1}^{|n|}(-i \cot(\epsilon))}{P_{\Delta -1}^{|n|}(+i \cot(\epsilon))P_{\Delta -1}^{|n|}(+\tanh (\tau_b ))-P_{\Delta -1}^{|n|}(-i \cot(\epsilon))P_{\Delta -1}^{|n|}(-\tanh (\tau_b ))}
\end{equation}
\begin{equation}
B_n=\frac{ \phi^b_n P_{\Delta -1}^{|n|}(+i \cot(\epsilon))-\epsilon^{\Delta} \phi_n P_{\Delta -1}^{|n|}(-\tanh (\tau_b ))}{P_{\Delta -1}^{|n|}(+i \cot(\epsilon))P_{\Delta -1}^{|n|}(+\tanh (\tau_b ))-P_{\Delta -1}^{|n|}(-i \cot(\epsilon))P_{\Delta -1}^{|n|}(-\tanh (\tau_b ))}
\end{equation}

With these coefficients at hand we can proceed to obtain the on shell action. The piece at the asymptotic boundary is
\begin{align}\label{ST}
S_T&=+\frac i2 \int_T\!\! \sqrt{\gamma} \;\Phi \,\partial_n \Phi-I_{\rm ct}
=\frac{1}{2}\sum_n \phi_{-n} \,\phi_n \psi_{2;n} +\frac{1}{2}
 \sum_n \phi_{-n} \,\phi^b_n K_{b;n}+\dots
\end{align}
where
\begin{equation}
\psi_{2;n}=-2\pi \frac{4^{\Delta } \Gamma \left(\Delta +\frac{1}{2}\right) \Gamma (1-|n|-\Delta) }{\Gamma \left(\frac{1}{2}-\Delta \right) \Gamma (\Delta -|n|) }\times\frac{ \left(P_{\Delta -1}^{|n|}(-\tanh (\tau_b))+e^{i \pi  (\Delta +|n|)} P_{\Delta -1}^{|n|}(+\tanh (\tau_b))\right)}{ \left(e^{i \pi  \Delta } P_{\Delta -1}^{|n|}(-\tanh (\tau_b))-e^{i \pi  |n|} P_{\Delta -1}^{|n|}(+\tanh (\tau_b))\right)}
\end{equation}
\begin{equation}
K_{b;n}=\pi\frac{\sqrt{\pi } 2^{\Delta +1} e^{\frac{1}{2} i \pi  (\Delta +|n|)}}{\Gamma \left(\frac{1}{2}-\Delta \right) \Gamma (\Delta -|n|)}\times \frac{1}{\left(e^{i \pi  \Delta } P_{\Delta
   -1}^{|n|}(-\tanh (\tau_b))-e^{i \pi  |n|} P_{\Delta -1}^{|n|}(+\tanh (\tau_b))\right)}
\end{equation}
where we have split the $\tau_b$ dependence. 
A Fourier transform into configuration space for these contributions appear to be beyond reach, but we can explore the $\tau_b\to-\infty$ limit. We can motivate this as an exploration of the states closest the Euclidean Vacuum. 

For the contributions to the on shell action found above we have,
\begin{align}
\frac{ \left(P_{\Delta -1}^{|n|}(-\tanh (\tau_b))+e^{i \pi  (\Delta +|n|)} P_{\Delta -1}^{|n|}(+\tanh (\tau_b))\right)}{ \left(e^{i \pi  \Delta } P_{\Delta -1}^{|n|}(-\tanh (\tau_b))-e^{i \pi  |n|} P_{\Delta -1}^{|n|}(+\tanh (\tau_b))\right)}&\sim \\
   & \hspace{-5cm} -e^{i \pi  \Delta }-e^{2 |n| \tau_b}\frac{\pi  \left(1+e^{2 i \pi  \Delta }\right) \csc (\pi 
   \Delta ) \Gamma (|n|+\Delta )}{\Gamma (|n|) \Gamma
   (|n|+1) \Gamma (\Delta -|n|)}+\dots
\end{align}
\begin{equation}
\frac{1}{\left(e^{i \pi  \Delta } P_{\Delta
   -1}^{|n|}(-\tanh (\tau_b))-e^{i \pi  |n|} P_{\Delta -1}^{|n|}(+\tanh (\tau_b))\right)} \sim e^{|n| \tau _b} \frac{\pi  e^{i \pi  \Delta } \csc (\pi  \Delta ) }{\Gamma (|n|)}+\dots
\end{equation}
Before computing correlation functions from these contributions, we compute $I_b$. This piece of the on shell action is finite on its own and yields,
\begin{align}\label{Sb}
S_b
&=   \sum_n \phi^{b}_{-n}\phi^b_n \left(-\pi \Delta \tanh \left(\tau _0\right)+ K^{bb}_n\right) +\frac{1}{2}
 \sum_n \phi_{-n} \,\phi^b_n K_{b;n}+\dots
\end{align}
with
\begin{equation}
K^{bb}_n=\pi(\Delta-|n| )\frac{\left(e^{-i \pi  \Delta } P_{\Delta }^{|n|}\left(+\tanh \left(\tau _0\right)\right)+e^{-i \pi  |n|} P_{\Delta }^{|n|}\left(-\tanh \left(\tau _0\right)\right)\right)}{ \left(e^{-i \pi 
   \Delta } P_{\Delta -1}^{|n|}\left(+\tanh \left(\tau _0\right)\right)-e^{-i \pi  |n|} P_{\Delta -1}^{|n|}\left(-\tanh \left(\tau _0\right)\right)\right)}\sim -\pi(\Delta-|n|)+\dots
\end{equation}

From our on shell action $S_{\rm fin}$ using \eqref{Sfin}, \eqref{ST} and \eqref{Sb}, we can compute the correlator functions for the boundary CFT at late times in the state defined by $\psi_b$. The two point function yields
\begin{align}\label{BP-2pf}
    \langle {\cal O}(\varphi') {\cal O}(\varphi) \rangle_{J_b} 
    =\langle {\cal O}(\varphi') {\cal O}(\varphi) \rangle_0\left(1+\frac{\langle {\cal O}(\varphi') {\cal O}(\varphi) \rangle_{\rm corr}}{\langle {\cal O}(\varphi') {\cal O}(\varphi) \rangle_0}+\dots\right)
\end{align}
where $\langle {\cal O}(\varphi') {\cal O}(\varphi) \rangle_0$ is the standard Euclidean vacuum 2 point function as defined in \eqref{NBP-2pf-varphi} and 
\begin{equation}\label{2pf-Corr}
    \langle {\cal O}(\varphi) {\cal O}(0) \rangle_{\rm corr}=\frac{e^{2\tau_b} \left(1+e^{2 i \pi  \Delta }\right)  \csc ^2(\pi  \Delta ) \Gamma \left(\Delta
   +\frac{1}{2}\right) }{\pi ^{-2} 2^{-2 \Delta -1}\Gamma
   \left(\frac{1}{2}-\Delta \right) \Gamma (\Delta -1)^2} \Re\left(e^{i \varphi } \, _2F_1\left(2-\Delta
   ,2-\Delta ;2;-e^{i \varphi +2 \tau _b}\right)\right)
\end{equation}
The one point function yields
\begin{equation}
   \langle {\cal O}(\varphi) \rangle_{J_b}= \int d\varphi \; \phi_b(\varphi) K_{b}(\varphi) + \dots
\end{equation}
with 
\begin{figure}[t]\centering
\begin{subfigure}{0.49\textwidth}\centering
\includegraphics[width=.9\linewidth] {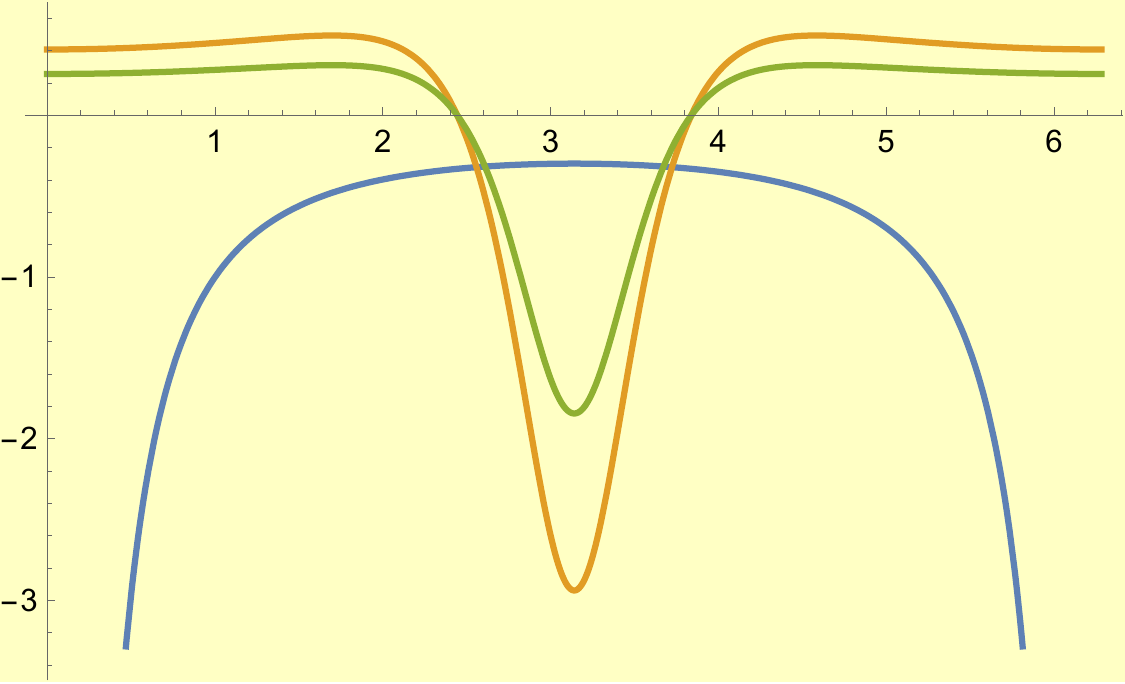}
\caption{}
\end{subfigure}
\begin{subfigure}{0.49\textwidth}\centering
\includegraphics[width=.9\linewidth] {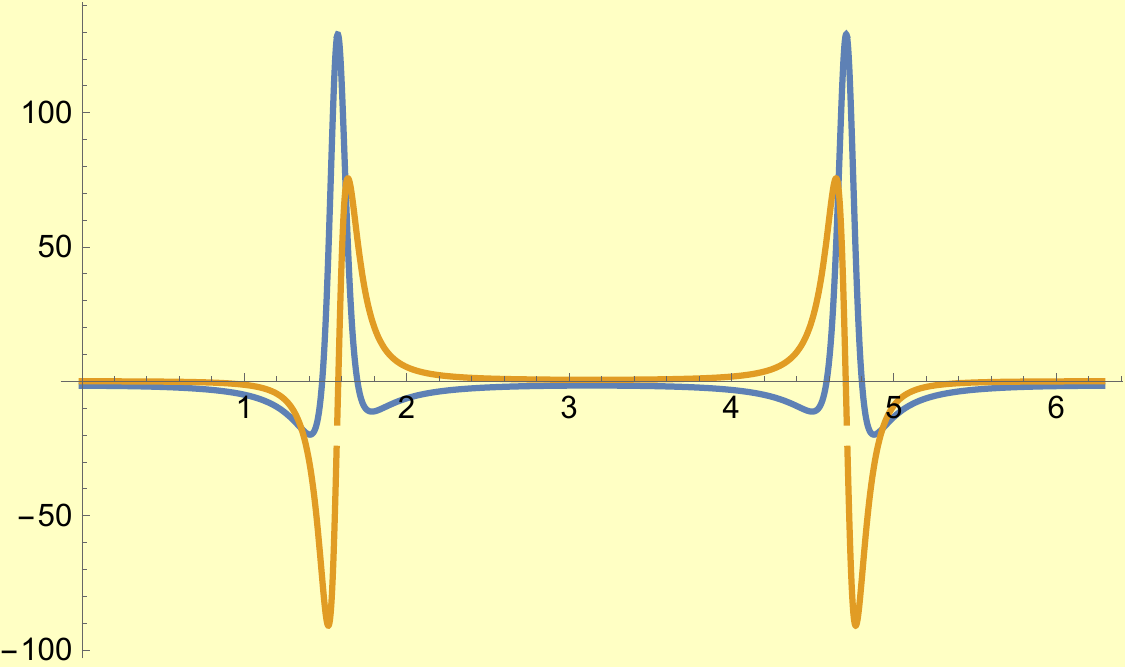}
\caption{}
\end{subfigure}
\caption{In (a) we show $\langle {\cal O}(\varphi') {\cal O}(\varphi) \rangle_0$ in blue and $\frac{\langle {\cal O}(\varphi') {\cal O}(\varphi) \rangle_{\rm corr}}{\langle {\cal O}(\varphi') {\cal O}(\varphi) \rangle_0}$ real and imaginary pieces in orange and green. In (b) we are showing a sample form for the real and imaginary piece of $K_b(\varphi)$}
\label{Fig:Corrs}
\end{figure}
\begin{equation}
    K_{b}(\varphi)=\frac{2 \pi ^{3/2} \csc (\pi  \Delta ) e^{-\frac{1}{2} i \pi  (\Delta
   -1)-\tau_b+i \varphi } }{\Gamma \left(\frac{1}{2}-\Delta \right) \Gamma (\Delta
   -1)} \left(\frac{\left(2+2 i e^{\tau_b-i \varphi }\right)^{\Delta }}{\left(e^{-\tau_b+i \varphi
   }+i\right)^2}+\frac{\left(2+2 i e^{\tau_b+i \varphi
   }\right)^{\Delta }}{\left(e^{-\tau_b}+i e^{i \varphi
   }\right)^2}\right)
\end{equation}
and the wavefunctional get renormalized by the $\phi_b$ as
\begin{equation}
    \ln Z^{L+E}_{dS;\rm ren}(\phi=0)\sim - \frac{1}{8\pi} \int d\varphi d\varphi' \phi_b(\varphi) \left(\frac{1}{\sin^{2}\left(\frac{\varphi-\varphi'}{2}\right)}\right) \phi_b(\varphi') + \dots
\end{equation}
We show sample profiles of $\langle {\cal O}(\varphi') {\cal O}(\varphi) \rangle_{\rm corr}$ and one point function kernel $K_b(\varphi)$ in Fig \ref{Fig:Corrs}.

We conclude this section by listing the salient aspects of this example. First, we have obtained that $\phi_b\to0$ recovers the $\langle{\cal O}({\bf x})\rangle_{J_b}\to0$ as well as the renormalization of the wavefunction $ Z^{L+E}_{dS;\rm ren}(\phi=0)\to1$. This is consistent with treating $\phi_b$ in the dual CFT as a source for the operator ${\cal O}$ dual to $\Phi$. We have also found the form of $K_b$ in eq. \eqref{1pt-exc1} to leading order in a small hole $\tau_b\to-\infty$ expansion. 
The 2 point function correction on its own is seen to be independent of $\phi_b$, which does not contradict its interpretation as $J_b$. The change is completely due to the boundary at $\tau_b$ being introduced. In particular, we also stress that the 2 point function \eqref{2pf-Corr} has no antipodal $\varphi=\pi$ divergencies as the correlators coming from $\alpha$-vacua as reviewed in \eqref{OO-CQ-varphi-alpha}.
We therefore also conclude that the states constructed by imposing the $\Sigma_b$ boundary cannot be related (at least pertubately) to $\alpha$-vacua in our theory. Thus, the correlators reached via non-trivial $J_b \neq 0 \Longleftrightarrow (\Sigma_b, \phi_b)$, describe a different family of states.

\section{Final Remarks and Future Directions}

In this article we proposed a generalization of the Hartle-Hawking construction in de Sitter spacetime. We argued that NBP might be relaxed, and so more general initial states can be prepared through a Euclidean path integral, which by construction, yields a good geometry and fields configuration as semiclassical limit. The standard holographic formulas were preserved, and moreover, it was shown that such generalization is consistent with the simplest notion of excitations in a CFT. Consequently, we found new formulas that connect the n-point functions in this theory with late-time correlators in de Sitter spacetime.

As a first approach, we neglected fluctuations involving gravity, which requires a more detailed and in-depth analysis of the GR constraints, such as the gravitational Gauss law \cite{Higuchi1,Higuchi2}. One question would be, for example, how the Wheeler-De Witt equation should be modified and whether the new states are solutions. This will be addressed in a future work.

On the other hand, discrepancies between cosmological observations and the slow-roll inflation model in the no-boundary geometry have been pointed out recently \cite{malda24}. Future research could explore the states constructed here in order to better fit these observations.

In addition, we also explored whether the proposed states make contact with $\alpha$-vacua, but we find no evidence for this. 
In \cite{Strom-alpha-vac,otro-alpha-vac} $\alpha$-vacua were studied in the context of dS/CFT, and were associated with a double-trace deformation using ${\cal O}_{\Delta}{\cal O}_{d-\Delta}$ which is always of dimension $d$. This is akin to the standard interpretation of multi-trace operator deformations in AdS/CFT  \cite{Witten:2001ua} in the sense that it is consistent with changing the boundary conditions of the field. This is also consistent with our results: no multi-trace operators were sourced and we found no trace of $\alpha$-vacua in the 2 point function \eqref{2pf-Corr}.

As a final note, the No-Boundary Proposal (NBP) was originally interpreted as describing the universe emerging from nothing, with the no-boundary geometry representing the amplitude for the spatial geometry $\Sigma$
 to arise from a single point rather than from a surface \cite{Vilenkin1,Vilenkin2}. However, from a holographic perspective, a more suitable interpretation is to regard the Euclidean geometry as the dual holographic description of the exact quantum state at the semiclassical level. Thereby, the boundary shall represent the region where this description effectively breaks down.

\vspace{.5cm}
\textbf{Acknowledgements:} 
We thank Guillermo Silva for helpful discussions. The authors are grateful to CONICET and UNLP for financial support. FLC is grateful to the Universidad Adolfo Ibáñez which hosted him in the latter stages of this project.

\pagebreak

\appendix

\section{Holographic Correlators in dS in the NBP}\label{NBPCorrelators}

In this section we cover the standard computation of holographic CFT correlators in the Euclidean Vacuum of de Sitter. This state is often defined in Poincaré foliation and is also called Bunch-Davies vacuum \cite{BDVacuum}, but for the purposes of showing the mixed signature techniques and fixing notation for Sec. \ref{BPCorrelators} we cover the computation in detail. 
The geometry under study is presented in Fig. \ref{NBP1}. We foliate the manifold as in conformal coordinates as
\begin{equation}\label{Euclid-Metric-NBP}
ds_{dS}^2=\frac{d\tau^2+d\varphi^2}{\cosh(\tau)^2}\qquad\qquad\tau\in(-\infty,0] \qquad\varphi\in[0,2\pi)
\end{equation}
where we defined the south pole at $\tau\to-\infty$. We go to the Lorentzian signature piece as 
\begin{equation}\label{Lorentz-Metric-NBP}
\tau\to -i t \qquad\qquad\cosh(\tau)\to\;\cos(t)\qquad\qquad ds_{dS}^2\;\to\;\frac{-dt^2+d\varphi^2}{\cos(t)^2}\qquad t\in[0,\pi/2-\epsilon]
\end{equation}
with the same domain for $\varphi$ and $\epsilon\ll1$ being a cut-off to the late-time $I^+$ region sitting at $t=\pi/2$. The Euclidean and Lorentzian pieces are glued at $\tau=t=0$, where one can check that the standard junction conditions are met \cite{Israel}.

We want to compute the on shell action for a massive scalar on this geometry and compute late time correlators. The relations between the field and renormalized quantities involves the holographic dictionary $\Lambda$ defined as the factor that we remove from dS to define the CFT metric, i.e.
\begin{equation}\label{B-Metric-NBP}
ds_{CFT}^2=\lim_{\epsilon\to0}\Lambda^{-2}\left(ds_{dS}^2\right)_{t=\pi/2-\epsilon}=d\varphi^2 \qquad\Rightarrow\qquad \Lambda=\frac{1}{\cos(\pi/2-\epsilon)} \sim\frac{1}{\epsilon}
\end{equation}

As for the scalar field on the geometry, we define the Euclidean action as
\begin{align}\label{Bare-ScalarS-NBP}
S_E=\frac 12 \int\!\! \sqrt{g} \left(\partial_\mu\Phi\partial^\mu\Phi+m^2\Phi^2\right)=\frac 12 \int\!\! \sqrt{\gamma} \;\Phi \,\partial_n \Phi -\frac 12 \int\!\! \sqrt{g}\; \Phi\left(\square-m^2\right)\Phi
\end{align}
and the Lorentzian piece is straightforward to obtain from \eqref{Lorentz-Metric-NBP}. The full action contains both an Euclidean and a Lorentzian piece as is written in \eqref{wf0-onshell}.
The bare on-shell action for the free field can be written as a boundary term, the boundary contributions at $t=\tau=0$ cancel each other by virtue of the gluing conditions \eqref{gluing}, see \cite{SvR1,SvR2,Israel}, so that we get
\begin{equation}
S_{\rm ren}=S-S_{ct}=+\frac i2 \int_\epsilon\!\! \sqrt{\gamma} \;\Phi \,\partial_n \Phi-I_{ct}=+\frac i2 \int_\epsilon\!\! \;\Phi \,\partial_t \Phi-I_{ct}
\end{equation}
where we used that $\sqrt{\gamma}\partial_n=\partial_t$ in conformal coordinates, $S_{\rm ct}$ are counter-terms which we define below that render the on shell action finite. We now discuss in detail the properties of the solutions $\Phi$ to the classical solutions to the equations of motion and compute the correlators from the renormalized on-shell action defined on the manifold defined above in Fig. \ref{NBP1}. These are
\begin{equation}
\left(\square-m^2\right)\Phi=0\qquad\qquad\Phi_{n}(\pi/2-\epsilon)=\tilde\phi_n 
\end{equation}
where $\tilde\phi_n $ is a bare source that we will renormalize below ans we have defined Fourier transforms with the convention,
\begin{equation}
    \Phi_{n}(\tau)=\int_0^{2\pi} \frac{d\varphi}{2\pi} e^{-in\varphi}\Phi(\tau,\varphi) \qquad\qquad \Phi(\tau,\varphi)=\sum_{n\in\mathbb{Z}} e^{+in\varphi}\Phi_{n}(\tau)
\end{equation}
and is regular as $\tau\to-\infty$. This determines a unique the solution, which can be written as  
\begin{equation}
\Phi^E_{n}(\tau)=\tilde\phi_n \;\frac{P_{\Delta -1}^{|n|}(-\tanh (\tau ))}{P_{\Delta -1}^{|n|}(i \cot (\epsilon ))}\qquad\qquad
\Phi^L_{n}(t)\equiv \Phi^E_{n}(-i t)
\end{equation}
so that the bare on shell action is
\begin{align}
S&=+\frac i2 \int_\epsilon\!\! \;\Phi \,\partial_t \Phi= +i\pi \sum_n \tilde \phi_{-n}\tilde \phi_{n} \, \left(\partial_t\frac{P_{\Delta -1}^{|n|}(i\tan(t))}{P_{\Delta -1}^{|n|}(i\cot(\epsilon))}\right)_{t=\pi/2-\epsilon}\\
&=i\pi \sum_n \tilde \phi_{-n}\tilde \phi_{n}\left(\Delta  \cot (\epsilon )-i (\Delta-{|n|})\frac{ P_{\Delta }^{|n|}(i \cot (\epsilon
   ))}{P_{\Delta -1}^{|n|}(i \cot (\epsilon
   ))}\right)
\end{align}
Now, the bare on shell action is $\epsilon^{-1}$ divergent as expected from the analogous AdS/CFT computation, see e.g. \cite{HolRenormAdS,HolRenormdS}. We remove this divergence via a counterterm,
\begin{align}\label{CT-Action}
I_{\rm ct}&=i\pi\Delta \sum_n \tilde \phi_{-n}\tilde \phi_{n}\left(\sqrt{\gamma}-\frac\epsilon2\right)
=\frac i2 \Delta \int \sqrt{\gamma} \;\Phi^2 +\dots \qquad\quad \cot (\epsilon)=\sqrt{\gamma}-\epsilon/2
\end{align}
The second piece contains the information of the correlator. As with the metric, the sources $\tilde\phi_{n}$ need to be renormalized according to the dual operator's conformal dimension $\Delta$. Again, in analogy with AdS/CFT, from an asymptotic analysis at $t= \pi/2-\epsilon$ for $0<\epsilon\ll1$ of the classical solution,
\begin{equation}
P_{\Delta -1}^{|n|}(-i\cot(\epsilon))\sim
\epsilon ^{1-\Delta } \frac{i
   2^{\Delta -1}  \Gamma \left(\Delta
   -\frac{1}{2}\right) 
   }{\sqrt{\pi } \Gamma (\Delta
   -{|n|})e^{\frac{1}{2} i \pi 
   ({|n|}+\Delta )}}
   +\epsilon ^{\Delta }
   \frac{2^{-\Delta }
   \Gamma \left(\frac{1}{2}-\Delta
   \right)}{\sqrt{\pi } \Gamma (1-{|n|}-\Delta)e^{\frac{1}{2} i \pi  ({|n|}-\Delta )}}+\dots
\end{equation}
\cite{HolRenormAdS}
where due to our choice $\Delta\in[0,1/2]$, we see that the leading behavior is $\epsilon^{\Delta}$. From this we read that the dual operator dimension is $1-\Delta$ and that the renormalized source $\phi_{n}$ is \footnote{At this point the Complementary representation choice simplifies drastically the analysis with respect to, e.g. the Principal Series in which $\Delta\in\mathbb{C}$. In such scenarios the dS/CFT the holographic map is more subtle, see e.g. \cite{Gizem1}.}
\begin{equation}\label{Source-Renorm}
    \tilde\phi_n \equiv \Lambda^{-\Delta} \phi_{n} \sim \epsilon^{\Delta}\phi_{n}
\end{equation}
With this rescaling of the sources, the finite piece of the on shell action yields,
\begin{align}
S_{\rm ren}&= \sum_n \phi_{-n}\phi_{n} \left(\pi\epsilon^{2\Delta} 
   (\Delta-|n| )\frac{P_{\Delta }^{|n|}(i \cot (\epsilon
   ))}{P_{\Delta -1}^{|n|}(i \cot (\epsilon
   ))}\right)\\
 &  = 4^{\Delta }\pi e^{-i \pi 
   \Delta }\frac{ \Gamma \left(\Delta
   +\frac{1}{2}\right)}{\Gamma
   \left(\frac{1}{2}-\Delta
   \right)}\sum_n \phi_{-n}\phi_{n} \frac{\Gamma
   (1-|n|-\Delta)}{\Gamma (\Delta -|n|)}\\
&=\frac{1}{2}\int d\varphi d\varphi' \phi(\varphi) \left(\frac{2^{4\Delta -3}  \csc (\pi  \Delta )
   \Gamma \left(\Delta
   +\frac{1}{2}\right)}{e^{i \pi 
   \Delta }\Gamma
   \left(\frac{1}{2}-\Delta
   \right) \Gamma (2 \Delta -1)}\frac{1}{\sin \left(\frac{\varphi-\varphi'
   }{2}\right)^{2(1-\Delta)} }\right)\phi(\varphi')
\end{align}
where we used the result
\begin{equation}
    \sum_{n\in\mathbb{Z}}e^{i n \varphi}\frac{\Gamma
   (1-|n|-\Delta)}{\Gamma (\Delta -|n|)}=\frac{\pi  4^{\Delta -1} \csc (\pi
    \Delta ) }{\Gamma
   (2 \Delta -1)}\frac{1}{\sin \left(\frac{\varphi
   }{2}\right)^{2(1-\Delta)} }
\end{equation}
This last expression implies
\begin{align}\label{NBP-2pf-varphi}
\langle {\cal O}(\varphi') {\cal O}(\varphi) \rangle_0 & 
=-\frac{\delta S_{\rm ren}(\phi)}{\delta \phi(\varphi')\delta \phi(\varphi)}=-\frac{2^{4\Delta -3}  \csc (\pi  \Delta )
   \Gamma \left(\Delta
   +\frac{1}{2}\right)}{e^{i \pi 
   \Delta }\Gamma
   \left(\frac{1}{2}-\Delta
   \right) \Gamma (2 \Delta -1)}\frac{1}{\sin^{2(1-\Delta)} \left(\frac{\varphi-\varphi'
   }{2}\right) }
\end{align}
and
\begin{align}\label{NBP-2pf-n}
\langle {\cal O}_n {\cal O}_m \rangle_0 & \equiv \frac{\delta \ln Z^{L+E}_{dS;\rm ren}}{\delta \phi^r_n\delta \phi^r_m}\sim-\frac{\delta S_{\rm ren}}{\delta \phi^r_n\delta \phi^r_m}= \delta_{n+m} (2\pi) 4^{\Delta } e^{-i \pi 
   \Delta }\frac{ \Gamma \left(\Delta
   +\frac{1}{2}\right)}{\Gamma
   \left(\frac{1}{2}-\Delta
   \right)}\frac{\Gamma
   (1-|n|-\Delta)}{\Gamma (\Delta -|n|)}
\end{align}
where the subindex $0$ is notation for the Euclidean vacuum.

\section{Searching for traces of \texorpdfstring{$\alpha$-vacua}{alpha-vacua}}\label{App:alpha-vac}

We now perform an additional computation of late time correlators in the NBP set-up to set the ground to study the $\alpha$-vacua \cite{Allen} 
contribution to late time correlators. We want to contrast these to the correlator deformations we find for our candidate excited states in eq. \eqref{2pf-Corr}. We will argue that, at least in perturbation theory, our deformations are not of the form of the ones that $\alpha$-vacua states would produce.

To do so, and following \cite{Allen}, see also \cite{Zimo}, we follow a canonical quantization approach. This will allow to estimate the contribution to $\langle {\cal O}(\varphi') {\cal O}(\varphi) \rangle$ that would rise from an $\alpha$-vacua contribution.
The mathematics involved are quite similar to our example in App. \ref{NBPCorrelators} , but the reasoning is quite different.
One needs to solve the Klein Gordon equation in dS$_{1+1}$ for the same massive field in Complementary representation $\Delta\in[0,d/2]$, $d=1$ in conformal coordinates \eqref{Euclid-Metric-NBP}-\eqref{Lorentz-Metric-NBP}, i.e. in the geometry described by Fig. \ref{NBP1} and impose regularity in the Euclidean south pole. This is enough to define energy-eigenstates and canonically quantize the field in Euclidean vacuum. From there, one can compute de Sitter 2 point functions and late time correlators for any $\alpha$-vacuum following in \cite{Allen}. We show the relation between these, \eqref{NBP-2pf-n} and the correlator \eqref{BP-2pf} found in the main text.
 
We write our canonically quantized field as 
\begin{equation}\label{CQ-Scalar-varphi}
\hat \Phi_L(t,\varphi)=\sum_{n\in\mathbb{Z}} \left( a_n \;\psi_{\Delta,n}(t,\varphi)+h.c.\right)\qquad\qquad \psi_{\Delta,n}(t,\varphi) \equiv \mathbb{C}_{n}^\Delta \,e^{i n \varphi} P_{\Delta -1}^{|n|}(i \tan (t))
\end{equation}
\begin{equation}\label{CnDelta}
\mathbb{C}_{n}^\Delta=\sqrt{\frac{\Gamma (1-|n|-\Delta) \Gamma
   (\Delta -|n|)}{4\pi}}\in\mathbb{R}
\end{equation}
The resulting 2 point function in Euclidean vacuum is
\begin{equation}
\langle \,\hat \Phi \,\hat \Phi\,\rangle\equiv G(Z)=\frac{\Gamma (1-\Delta) \Gamma
   (\Delta )}{4\pi}\,_2F_1\left(\Delta,1-\Delta,1;\frac{1-Z}{2}\right)
\end{equation}
where $Z$ is the geodesic distance between 2 points. This function is divergent at the north pole of the sphere at $Z=-1$. In the conformal coordinates, for points e.g. in the Euclidean section, we have
\begin{equation}
Z=\text{sech}(\tau )\text{sech}(\tau') \cos (\varphi -\varphi')+\tanh (\tau ) \tanh (\tau')
\end{equation}
We can now compare the late-time correlators coming from these dS correlators and find a perfect agreement with the relations between vacuum late time correlators and holographic $\langle {\cal O}{\cal O}\rangle$ in momentum space \eqref{NBP-2pf-n} as shown in e.g. \cite{Malda03,Bauman}
\begin{align}\label{LT-CQ-n}
\langle \hat \Phi^{LT}_{n} \hat \Phi^{LT}_{n}\rangle\equiv\lim_{t\to\infty} \epsilon ^{-2 \Delta }\langle \hat \Phi_{L;n}(t) \hat \Phi_{L;n}(t)\rangle
\propto\langle {\cal O}_n {\cal O}_{-n} \rangle^{-1}
\end{align}
where we defined the Fourier modes as
\begin{equation}\label{CQ-Scalar-n}
\hat \Phi_{L;n}(t)\equiv\int \frac{d\varphi}{2\pi} e^{-i n \varphi}\hat \Phi_L(t,\varphi)
=\mathbb{C}_{n}^\Delta \left(a_n P_{\Delta -1}^{|n|}(i \tan (t))
+ h.c.\right)
\end{equation}

With these results at hand, we can finally consider the potential contribution from the so called $\alpha$-vacua to the late time correlators $\langle \hat \Phi^{LT} \hat \Phi^{LT}\rangle$ and holographic correlators $\langle {\cal O} {\cal O} \rangle$. We can do this by noticing that our mode decomposition is exactly of the form of Allen-Motolla in the sense that $\psi_{\Delta,n}(t,\varphi)^*=\psi_{\Delta,n}(\bar t,\bar \varphi)$, where $\bar x$ is the antipodal point of $x$ in de Sitter, as defined in \cite{Allen}.  Using these modes, the correlator in any $\alpha$-vacua is seen to be
\begin{align}
\langle \,\hat \Phi \, \hat \Phi \,\rangle_\alpha
   &\equiv G_{\alpha}=\cosh^2(\alpha)G(Z)+\sinh^2(\alpha)G(\bar Z)
\end{align}
where $\bar Z(x,y)=-Z(x,y)$ is the chordal distance between $x$ and $\bar y$, the antipodal point of $y$. For $\alpha\neq 0$ one can see that $G_{\alpha}$ is both divergent on the north and south poles of the Euclidean sphere $Z=\pm 1$. 

The late time correlators in momentum space are straightforwardly computed as
\begin{align}\label{LT-CQ-n-alpha}
\langle \hat \Phi^{LT}_{n} \hat \Phi^{LT}_{n}\rangle_{\alpha}
&\sim \frac{\Gamma (\Delta -|n|)}{\Gamma (1-\Delta-|n|)}
   \left(\cosh^2(\alpha)+e^{i |n| \pi}\sinh^2(\alpha)\right)
\end{align}
Following the relation \eqref{LT-CQ-n} valid for any dS vacuum, we conclude that, at least expanding perturbately in $\alpha\ll1$ to all orders, the holographic correlators should contain non-local antipodal divergencies at $\Delta\varphi=\pi$ as well as contact divergencies at $\Delta\phi=0$, i.e.
\begin{equation}\label{OO-CQ-n-alpha}
\langle {\cal O}_{n} {\cal O}_{-n}\rangle_{\alpha}\sim \langle \hat \Phi^{LT}_{n} \hat \Phi^{LT}_{n}\rangle_{\alpha}^{-1}
\sim 
   \frac{\Gamma(1-\Delta-|n|)}{\Gamma (\Delta -|n|)}\left((1-\alpha^2)-e^{i n \pi}\alpha^2+O[\alpha]^4\right)
\end{equation}
\begin{equation}
\begin{aligned}\label{OO-CQ-varphi-alpha}
\langle {\cal O}(\varphi) {\cal O}(0)\rangle_{\alpha}
\sim& 
    \frac{1-\alpha^2}{\sin^{2(1-\Delta)} \left(\frac{\varphi}{2}\right)}-\frac{\alpha^2}{\cos^{2(1-\Delta)} \left(\frac{\varphi}{2}\right)}+O[\alpha]^4\\
    \sim&(1-\alpha^2)\langle {\cal O}(\varphi) {\cal O}(0)\rangle_{0}+\alpha^2\langle {\cal O}(\varphi) {\cal O}(\pi)\rangle_{0}+O[\alpha]^4
\end{aligned}    
\end{equation}
Our modified correlators found in Sec. \ref{BPCorrelators} do not have this type of divergencies to any order in a perturbative expansion around $\tau_b\to-\infty$.
This is a strong suggestion that the states constructed according to a boundary proposal, have no contribution from the $\alpha$-vacua.

\end{document}